\documentclass[a4paper]{jpconf}
\usepackage{graphicx}
\usepackage{amssymb}
\newcommand{\OM}{\Omega_M}
\newcommand{\OMo}{\Omega_M^0}
\newcommand{\OXo}{\Omega_X^0}
\newcommand{\OL}{\Omega_{\Lambda}}

\newcommand{\rc}{\rho_c}

\newcommand{\OLo}{\Omega_{\Lambda}^0}
\newcommand{\rco}{\rho_{c}^0}
\newcommand{\mincir}{\raise
-3.truept\hbox{\rlap{\hbox{$\sim$}}\raise4.truept\hbox{$<$}\ }}
\newcommand{\magcir}{\raise
-3.truept\hbox{\rlap{\hbox{$\sim$}}\raise4.truept\hbox{$>$}\ }}
\newcommand{\be}{\begin{equation}}
\newcommand{\ee}{\end{equation}}
\newcommand{\rL}{\rho_{\Lambda}}
\newcommand{\rLi}{\rho_{\Lambda}^{i}}
\newcommand{\rLe}{\rho_{\Lambda{\rm eff}}}

\newcommand{\rF}{\rho_{F}}

\newcommand{\G}{{\cal G}}
\newcommand{\F}{{\cal F}}

\newcommand{\rLo}{\rho_{\Lambda}^0}
\newcommand{\CC}{\Lambda}
\newcommand{\rM}{\rho_M}
\newcommand{\pM}{p_M}
\newcommand{\rs}{\rho_{Ms}}

\newcommand{\weff}{\omega_{\rm eff}}
\newcommand{\wm}{\omega_{\rm m}}

\newcommand{\rD}{\rho_D}
\newcommand{\rX}{\rho_X}

\newcommand{\wX}{\omega_X}
\newcommand{\we}{\omega_e}

\newcommand{\tOM}{\tilde{\Omega}_M}

\newcommand{\tOD}{\tilde{\Omega}_{D}}
\newcommand{\drX}{\dot{\rho}_{\rm X}}
\newcommand{\zL}{\zeta_{\CC}}

\begin{document}
\title{Cosmologies with a time dependent vacuum}

\author{Joan Sol\`a}

\address{HEP Group, Dept. Estructura i Constituents de la Mat\`eria\\
and Institut de Ci\`encies del Cosmos,  Universitat de Barcelona,\\
Av. Diagonal 647, 08028 Barcelona, Catalonia, Spain}

\ead{sola@ecm.ub.es}

\begin{abstract}
The idea that the cosmological term $\CC$ should be a time dependent
quantity in cosmology is a most natural one. It is difficult to
conceive an expanding universe with a strictly constant vacuum
energy density, $\rL=\CC/(8\pi\,G)$, namely one that has remained
immutable since the origin of time. A smoothly evolving vacuum
energy density $\rL=\rL(\xi(t))$ that inherits its time-dependence
from cosmological functions $\xi=\xi(t)$, such as the Hubble rate
$H(t)$ or the scale factor $a(t)$, is not only a qualitatively more
plausible and intuitive idea, but is also suggested by fundamental
physics, in particular by quantum field theory (QFT) in curved
space-time. To implement this notion, is not strictly necessary to
resort to ad hoc scalar fields, as usually done in the literature
(e.g. in quintessence formulations and the like). A ``running''
$\CC$ term can be expected on very similar grounds as one expects
(and observes) the running of couplings and masses with a physical
energy scale in QFT. Furthermore, the experimental evidence that the
equation of state (EOS) of the dark energy (DE) could be evolving
with time/redshift (including the possibility that it might
currently behave phantom-like) suggests that a time-variable
$\CC=\CC(t)$ term (possibly accompanied by a variable Newton's
gravitational coupling too, $G=G(t)$) could account in a natural way
for all these features. Remarkably enough, a class of these models
(the ``new cosmon'') could even be the clue for solving the old
cosmological constant problem, including the coincidence problem.
\end{abstract}

\section{Introduction}
\vspace{0.5cm} The observed accelerated expansion of the
universe\,\cite{SNIa,WMAP} is currently one of the central issues of
observational and theoretical cosmology. It somehow came as a
surprise to discover that the universe is not progressively slowing
down its pace, but actually rocketing it up. The usual explanatory
paradigm for this striking observation assumes the existence of the
so-called {\em dark energy} (DE)-- a mysterious cosmic component
with negative pressure, which, in contrast to dark matter (DM), does
not cluster (at least in a comparable way) and pervades all corners
of the universe. An obvious candidate for the DE is a strictly
constant cosmological term, $\CC$. Nevertheless, this option
precludes the possibility of a time evolution of the DE (which might
be appealing so as to face the so-called coincidence problem, namely
the fact that the current matter density $\rM$ is so close to the DE
density $\rL=\CC/(8\pi G)$, or, equivalently, the fact that the
acceleration epoch started so recently). This drawback could be
cured by admitting that the CC can be a dynamical quantity:
$\CC=\CC(t)$, an option which is perfectly admissible by the
Cosmological Principle.

Actually, the mere possibility of a variable CC approach was
explored long ago on purely phenomenological grounds --
cf.\,\cite{oldvarCC1,overduin98} -- although many of the models are
little motivated. At present, we know that the idea of a slowly
running CC is in general compatible with the experimental
data\,\cite{BPS09a} and there are reasonable theoretical models
grounded on fundamental physical principles that can support these
ideas, meaning that there is no strict need to invoke ad hoc scalar
fields to generate a time dependence for the DE --
see\,\cite{ShapSol09} and references therein, see also
\cite{BFLWard10,Hindmarsh:2011hx} for recent alternative approaches
reinforcing this point of view. Still, the variable $\CC=\CC(t)$
models can be dealt with in terms of a self-conserved DE fluid in
the manner of a scalar field (as we cannot know a priori the very
nature of the DE, and hence we may be naturally tempted to keep on
using this simple scalar field approach). However, although this
description in terms of scalar fields (which may be called the ``DE
picture'' of the original $\CC=\CC(t)$ model\,\cite{SS123}) is
perfectly possible, the corresponding effective equation of state
(EOS) for the DE, $\omega=p_{\rm DE}/\rho_{\rm DE}$, will be a
non-trivial function of time or of the redshift, $\omega=\omega(z)$,
and in general it cannot be reduced to the standard forms proposed
in the literature. In other words, the effect of having a variable
cosmological term $\CC=\CC(t)$ cannot be described by the simple
parameterizations usually employed for $\omega(z)$, characterized by
two parameters $(\omega_0,\omega_1)$ that become constrained by the
observations\,\cite{EOSquint}. The effective EOS of a variable
vacuum model is in general more complicated than that. This is shown
in detail, with specific examples, in reference \cite{SS123}. In
particular, the vacuum models that we are going to consider cannot
be described with these simple parameterizations. Therefore, the
$\CC=\CC(t)$ models must be studied on their own and constitute an
independent class of DE models. It means that we should be more open
minded on the form that the effective EOS can adopt, and we should
not limit ourselves to some admittedly popular (albeit perhaps too
simple) forms utilized too often in the literature.

Remarkably, the variable $\CC=\CC(t)$ approach can also be
advantageously employed for solving (or highly palliating) the big
or ``old'' CC problem\,\cite{CCproblem}, viz.\, the absolutely
formidable task of trying to explain the value itself of the current
vacuum energy density $\rLo\sim 10^{-47}$ GeV$^4$ on the face of its
enormous input value $\rLi\sim M_X^4$ (with $M_X\sim 10^{16}$ GeV)
left over in the very early times, presumably just after the
inflation phase induced by a Grand Unified Theory (GUT). This option
has been recently investigated in\,\cite{BSS0910,BSSFull10a}, and we
will also review it here. It is based on the idea of ``relaxation''
of the effective vacuum energy density $\rL(t)=\CC(t)/(8\pi G)$.
Relaxation offers indeed an entirely different perspective on the
old CC problem; in particular, it provides the extremely appealing
possibility to solve the CC problem \textit{dynamically}, hence
\textit{without fine tuning}\,\cite{BSSFull10a}. Moreover, the
current value of the expansion rate $H$ becomes determined by an
approximate relation of the form $H_0\sim 1/(\rLi)^n\ (n>0)$.
Interestingly enough, this relation tells us that $H_0$ is small
just because the initial vacuum energy $\rLi$ was very large!

In the next sections, I summarize the general notion and properties
of the models with time-dependent vacuum energy, and consider
specific scenarios possessing very interesting features concerning
the possible solution, or at least significant alleviation, of some
of the most fundamental cosmological constant problems of modern
times.

\section{Two cosmological pictures: ``CC picture'' versus ``DE picture''}
\label{CCDEpicture}

\vspace{0.5cm} Suppose that we are given a model where the vacuum
energy density $\rL=\rL(\xi(t))$ is time evolving through its
dependence on other cosmological function(s) of the cosmic time,
e.g. $\xi(t)=\rM(t),H(t),a(t),...$. We take this starting
``picture'' as the original (physical) representation of the DE, and
we call it ``the CC picture''. In general, other fundamental
``constants'' can also be varying, in particular the gravitational
coupling $G$. Therefore, in the CC picture we will have in general
two functions, at least, of the cosmic time or the cosmological
redshift, namely
\begin{equation}\label{varibleCCG}
\rL(z)=\rL(\rM(z),H(z),...)\,,\ \ \ \ \ G(z)=G(\rM(z),H(z),...)\,.
\end{equation}
Of course other fundamental parameters could also be variable. For
example, the fine structure constant has long been speculated as
being potentially variable with the redshift. However, here we focus
purely on the potential variability of the most genuine fundamental
gravitational parameters of Einstein's equations, to wit: $\CC$ and
$G$.

For a cosmological model of the kind (\ref{varibleCCG}), the general
Bianchi identity of the Einstein tensor leads to
$\label{eq:covBianchi} \nabla^{\mu} \left[G (T_{\mu \nu} + g_{\mu
\nu} \rho_{\Lambda}) \right]=0$. In the FLRW metric, and modeling
the universe as a perfect fluid, it implies
\begin{equation}\label{BianchiGeneral}
\frac{d}{dt}\,\left[G(\rM+\rho_{\Lambda})\right]+3\,G\,H_{\rm
CC}\,(\rM+\pM)=0\,.
\end{equation}
This ``mixed" conservation law connects the variation of
$\rho_{\Lambda}$ and $G$ with that of the matter density $\rM$. As a
result, the matter fluid in the CC picture is generally
non-conserved and its evolution may be non-canonical, as we will see
later on in an specific example. The general expression for the
Hubble function in the CC picture, assuming a spatially flat FLRW
universe, is
\begin{equation}\label{HLambda}
H^2_{\rm CC}(z)=\frac{8\pi G(z)}{3}(\rM(z)+\rL(z))\equiv
H^2_0\,\left[\Omega_{M}^0\,f_{M}(z)(1+z)^{ \alpha}
+\Omega_{\Lambda}^0\,f_{\Lambda}(z)\right]\,,
\end{equation}
where $\alpha=3(1+\wm)$, with $\wm=0, 1/3\ $ for the matter and
radiation epochs respectively. Here $f_{M}$ and $f_{\Lambda}$ are
certain functions of the redshift. They must satisfy $f_{M}(0)=1$
and $f_{\Lambda}(0)=1$, in accordance with the current cosmic sum
rule $\Omega_{M}^{0}+\Omega_{\Lambda}^{0}=1$ for a flat universe.

On the other hand, the above cosmological dynamics in the CC picture
can also be described in terms of an alternative fluid with
covariantly self-conserved densities $(\rs,\rD)$, therefore
satisfying $\dot{\rho}_{Ms}+\alpha\,H_{\rm DE}\,\rs=0$ and
$\dot{\rho}_D+3\,H_{\rm DE}\,(1+\weff)\rD=0$, where $\weff=p_D/\rD$
will be, in general, a function of the redshift: $\weff=\weff(z)$.
Of course $\rs\neq\rM$ and $\rD\neq\rL$ in general. The Hubble rate
in the new picture takes on the form:
\begin{equation}\label{DEpicture}
H_{\rm DE}^2=\frac{8\pi
G_0}{3}(\rs(z)+\rD(z))=H^2_0\,\left[\tOM^0\,(1+z)^{\alpha}
+\tOD(z)\right]\,,
\end{equation}
where
\begin{equation}\label{rhoD}
\tOD(z)=\tOD^0\,\exp\left\{3\,\int_0^z\,dz'
\frac{1+\weff(z')}{1+z'}\right\}\,.
\end{equation}
We have defined  $\tOM^0=\rs^0/\rc^0$, with $\rc^0=3H_0^2/(8\pi
G_0)$ the current value of the critical density. Notice that the
parameter $\tOM^0$ need not coincide with $\OM^0$, as they are
determined from two different parameterizations of the DE data.
Similarly, we have defined $\tOD(z)=\rD(z)/\rc^0$ -- and its current
value $\tOD^0=\tOD(0)$. Notice also that $G_0=1/M_P^2$ ($M_P$ being
the Planck mass) is strictly constant, whereas $G$ in
eq.\,(\ref{HLambda}) is in general a variable function of the
redshift: $G=G(z)$. The cosmological picture characterized by the
\textit{self-conserved} pair of energy densities $(\rs,\rD)$ at
fixed $G_0$ will be called the ``DE picture''\,\cite{SS123}.

Since the two pictures are assumed to be equivalent descriptions of
the same cosmological evolution, their matching requires that the
expansion history of the universe is the same in both pictures, i.e.
that their Hubble functions are numerically equal:
\begin{equation}\label{matching}
H_{\rm DE}(z)=H_{\rm CC}(z)\,.
\end{equation}
Recall that we take the original picture (\ref{varibleCCG}) as the
physical one, whereas the DE picture plays the role of an effective
description of the former in terms of the self-conserved DE
component $\rD$ at fixed $G_0$. Therefore, the function $\weff(z)$
defined above can be considered as the ``effective EOS'' (in the
``DE picture'') of the original model (\ref{varibleCCG}). A
straightforward calculation, using the matching condition
(\ref{matching}) and the general Bianchi identity
(\ref{BianchiGeneral}), allows to determine the effective EOS
function $\weff(z)$ in terms of the fundamental parameters of the
original theory (\ref{varibleCCG}), as follows\,\cite{SS123}:
\begin{equation}\label{we2}
\weff(z)=-1+\frac13\,\frac{1+z}{\rD}\,\frac{d\rD}{dz}=-1+\frac{\alpha}{3}\,\left(1-\frac{\zL(z)}{\rho_D(z)}\right)\,,
\end{equation}
where we have defined $\zL(z)\equiv(G(z)/G_0)\,\rho_{\Lambda}(z)$,
and
\begin{equation}\label{IF}
\rD(z)=\left(1+z\right)^{\alpha}\left[\rD(0)-\alpha
\int_0^z\frac{dz'\,\zL(z')}{(1+z')^{(\alpha+1)}}\right]\,.
\end{equation}
In particular, for a model of the kind (\ref{varibleCCG}) with fixed
$G=G_0$ in the matter epoch ($\alpha=3$), the effective EOS simply
reads $\weff(z)=-\rL(z)/\rD(z)$. The latter would shrink to just
$\weff=-1$ only in case of a truly cosmological constant
$\rL=const.$ because then $\rD$ would coincide with it. With the
help of the matching condition (\ref{matching}) of the two pictures,
and the Bianchi identity, one can also show, with some effort, that
\begin{equation}\label{dzeta}
\frac{d\rD(z)}{dz}={\alpha\,(1+z)^{\alpha-1}}\,\rco\,
\left(\OM^0\,f_M(z)-\tOM^0\right)\,.
\end{equation}
This is a remarkable relation, if we recall the conditions
$f_i(0)=1$ fulfilled by the functions $f_M$ and $f_{\CC}$ in
eq.\,(\ref{HLambda}). Why is remarkable? As $\OM^0$ and $\tOM^0$
must be very close, and $f_M(z)$ is presumably monotonous,
eq.\,(\ref{dzeta}) shows that a value $z=z^*$ {\em always} exists
for which the \textit{l.h.s.} of eq.\,(\ref{dzeta}) vanishes and,
therefore, the effective EOS $\weff(z)$ crosses the phantom divide
at that point: $\weff(z^*)=-1$. Whether the crossing is in the
recent past or in the immediate future, it will depend on the
details of the original model (\ref{varibleCCG}). Notice, however,
that if $\tOM^0=\OM^0$, then the crossing will be exactly at
$z^*=0$, i.e. now. This shows that a generic model
(\ref{varibleCCG}) with time varying vacuum energy, when described
in the DE picture (i.e. as if it were a cosmological self-conserved
DE fluid) generally leads to a crossing of the phantom divide, thus
providing a possible natural explanation for the observational data
-- which still admit a tilt in the phantom domain\,\cite{WMAP}.

\section{Time dependent vacuum in the running $\CC$CDM model}
\label{RGmodel1}

In contradistinction to purely phenomenological models of time
dependent $\CC$\,\cite{oldvarCC1,overduin98}, we adopt here the
fundamental point of view that the $\CC$ term is a running parameter
in QFT in curved space-time\,\cite{Parker09}. As suggested in
\cite{ShapSol09}, we should have a corresponding renormalization
group (RG) equation of the general form
\begin{equation}\label{RGEG1a}
(4\,\pi)^2\,\frac{d\rL}{d\ln\mu}=
\sum_{n=1}^{\infty}\,A_n\,\mu^{2n}\,,
\end{equation}
in which $\mu$ is an arbitrary scale associated to the RG running.
Of course the full effective action is perfectly scale independent
(i.e. RG invariant)\cite{ShapSol09}, and the running of the vacuum
energy should ultimately reflect the dependence of the leading
quantum effects with respect to some physical cosmological quantity
$\xi=\xi(t)$ associated with $\mu$, hence $\rL=\rL(\xi)$. Despite it
being a matter of debate, it has been argued by different
methods\,\cite{SSS04,Fossil07} that the physical scale $\xi$ pointed
by the $\mu$-dependence is the Hubble rate $H$. Since $H=H(t)$
evolves with the cosmic time, the cosmological term inherits a
time-dependence through its primary scale evolution with $H$. The
coefficients $A_{n}$ in eq.\,(\ref{RGEG1a}) are obtained after
summing over the loop contributions of fields of different masses
$M_i$ and spins $\sigma_i$. The general behavior is $A_n\sim \sum
M_i^{4-2n}$\,\cite{SSS04,Fossil07}. Therefore, for $\mu=H\ll M_i$,
the series above is (for $n>1$) an expansion involving powers of the
small quantities $\mu=H$ and $H/M_i$. Given, however, that
$A_{1}\sim\sum M_i^2$, we see that the heaviest fields furnish the
dominant contribution. This trait (``soft-decoupling'') represents a
generalization of the decoupling theorem in QFT. In addition, being
$H_0\sim 10^{-33}\,eV$, the condition $\mu\ll M_i$ is amply met for
all known particles, and the series on the \textit{r.h.s} of
eq.\,(\ref{RGEG1a}) converges extremely fast at present. Also
important is the fact that only even powers of $\mu=H$ are
consistent with general covariance, see\,\cite{ShapSol09}. Finally,
the $n=0$ contribution is absent because it corresponds to terms
$\propto M_i^4$ that give an extremely fast evolution. Actually,
from the RG point of view they are already excluded because, as
noted above, $\mu\ll M_i$ for all known masses. In practice only the
first term $n=1$ is needed, with $M_i$ of the order of the highest
mass available.

As the dominant masses $M_i$ will be of order of a GUT mass scale
$M_X$ near the Planck mass $M_P$, it is convenient to introduce the
ratio
$\nu=\sigma/({12\pi})\,\left({M_X^2}/{M_P^2}\right)$,
in which $\sigma=\pm 1$ depending on whether bosons or fermions
dominate in their loop contributions to (\ref{RGEG1a}). If the
effective mass $M_X$ of the heavy particles is just $M_P$, the
parameter $\nu$ takes the value $\sigma\,\nu_0$, with $\nu_0\equiv
{1}/({12\,\pi})\simeq  2.6\times 10^{-2}$. In general we expect that
$\nu$ will take values of this order or below. Under the very good
approximation $n=1$, and with $\mu=H$, eq.\,(\ref{RGEG1a}) abridges
to
\begin{equation}\label{RGEG1b}
\frac{d\rL}{d\ln H}=\frac{3\,\nu}{4\,\pi}\,M_P^2\,H^2\,,
\end{equation}
whose solution is\,\footnote{It is interesting to notice that this
quadratic evolution law for the vacuum energy with the expansion
rate has also been suggested recently by alternative QFT methods,
see \cite{Maggiore10}.}
\begin{equation}\label{CCH}
\rL=c_0+c_1\,H^2\,,\\
\end{equation}
with
\begin{equation}\label{C0C1}
c_0=\rLo-\frac{3\,\nu}{8\pi}M_P^2\,H_0^2\,, \ \ \
c_1=\frac{3\,\nu}{8\pi}\,M_P^2\,.
\end{equation}
\begin{figure}[t]
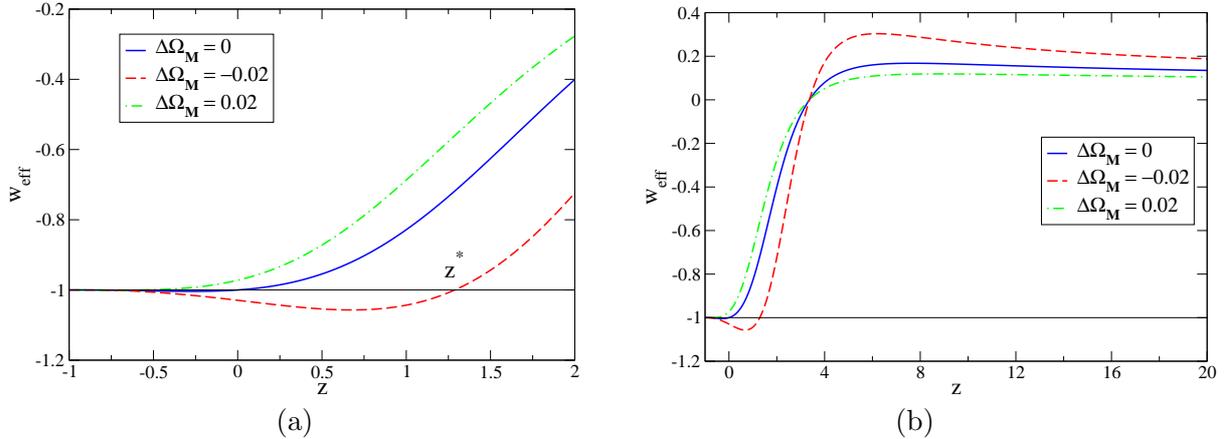

    \begin{tabular}{cc}
      \resizebox{0.47\textwidth}{!}{\includegraphics{fig1.eps}} &
      \hspace{0.3cm}
      \resizebox{0.47\textwidth}{!}{\includegraphics{fig2.eps}} \\
      (a) & (b)
    \end{tabular}
    \caption{Time dependent vacuum model (\ref{CCH}). \textbf{(a)} The effective EOS in the DE picture,
given by eq.\,(\protect\ref{wpflat1}), as a function of $z$ for
various $\Delta\Omega_M$ at fixed $\nu=-\nu_0$, and
$\Omega_M^0=0.3\,,\Omega_{\Lambda}^0=0.7$; \textbf{(b)} Extended $z$
range of the plot (a). $z^{*}$ is the crossing point of the phantom
divide $\weff=-1$.}
  \label{plot1}
\end{figure}
Solving the model in the original CC picture\,\cite{SS123} leads to
a non-standard evolution law for the matter density as a function of
the redshift: $\rM(z) =\rM^0\,(1+z)^{3(1-\nu)}$, along with the
following evolution of the vacuum energy density:
\begin{equation}\label{CRG}
\rL(z)=\rL^0+\frac{\nu\,\rM^0}{1-\nu}\,\left[(1+z)^{3(1-\nu)}-1\right]\,.
\end{equation}
For $\nu=0$ we recover the $\CC$CDM results, of course. But for
$\nu\neq 0$ we have a model in which the vacuum energy decays into
matter (or vice versa) according to the generalized conservation law
(\ref{BianchiGeneral}), with $G=const$ in this case. With this RG
model of the cosmic evolution at hand, we may now use it to apply
the procedure of section \ref{CCDEpicture} for obtaining the
effective properties of the model in the ``DE picture'' (in which
matter and dark energy are separately conserved). Obviously, for
this particular model, $\zL(z)=\rho_{\Lambda}(z)$. The corresponding
effective EOS cannot be described with standard
parameterizations\,\cite{EOSquint}. Indeed, from equation
(\ref{we2}) -- using (\ref{IF}) and (\ref{CRG}) -- we find:
\begin{equation}\label{wpflat1}
\hspace{-2cm} \weff(z)\left|_{\Delta\Omega\neq 0}\right.
=-1+(1-\nu)\,\frac{\Omega_M^0\,(1+z)^{3(1-\nu)}-\tilde{\Omega}_M^0\,(1+z)^3}
{\Omega_M^0\,[(1+z)^{3(1-\nu)}-1]-(1-\nu)\,[\tilde{\Omega}_M^0\,(1+z)^3-1]}\,.
\end{equation}
Here $\Delta\Omega_{M}\equiv \Omega_{M}^0-\tilde{\Omega}_{M}^0\neq
0$ is the difference of the cosmological mass parameters in the two
pictures (corresponding to two different fits of the same data). For
$|\nu|\ll 1$ we may expand the previous result to first order in
$\nu$. Assuming for simplicity that $\Delta\Omega_M=0$, we find
\begin{equation}\label{expwq}
\weff(z)\simeq-1-3\,\nu\frac{\Omega_M^0}{\Omega_{\Lambda}^0}\,(1+z)^3\,\ln(1+z)\,.
\end{equation}
This result reflects the essential qualitative features of the
analysis presented in the previous section, where in this case
$z^{*}=0$. Indeed, for $\nu>0$ eq.\,(\ref{expwq}) shows that we get
an (effective) phantom-like behavior ($\weff\lesssim-1$), whereas
for $\nu<0$ we have an (effective) quintessence behavior
($\weff\gtrsim-1$). In other words, this variable CC model can give
rise to two types of very different qualitative behaviors by
changing the sign of a single parameter. In Fig.\,\ref{plot1} we
show also more general cases where $z^{*}\gtrsim 0$, corresponding
to $\Delta\Omega_M\neq 0$. In contrast to the previous situation,
here the variable CC model may exhibit phantom behavior for $\nu<0$
(if $\Delta\Omega_M<0$), and manifests itself through the existence
of a transition point $z^{*}$ in our recent past  -- marked
explicitly in the figure -- as predicted by the general formalism of
section \ref{CCDEpicture}.

Let us also examine another interesting situation. From
eq.\,(\ref{BianchiGeneral}), we see that if $G=G(t)$ is variable and
we assume that matter is conserved (thus no transfer of energy with
the time dependent vacuum $\rL$), we have
${d\zL}=-(\rM/G_0)\,{dG}=(\rM^0/G_0)(1+z)^{\alpha}\,dG$. From
eq.\,(\ref{IF}) it is then not difficult to show that\footnote{ One
first shows that $\rD$ in eq.\,(\ref{IF}) satisfies the differential
equation
${d\rD(z)}/{dz}=\alpha\,\left(\rD(z)-\zL(z)\right)/{(1+z)}$. Next
one notes that the previous equation is also solved by the
alternative expression below:
$$
\rD(z)=\zL(z)-\left(1+z\right)^{\alpha}\,
\int_{z^{*}}^z\frac{dz'}{(1+z')^{\alpha}}\frac{d\zL(z')}{dz'}\ \ \
\Rightarrow \ \ \
\frac{d\rD(z)}{dz}=-\alpha\,\left(1+z\right)^{\alpha-1}
\int_{z^{*}}^z\frac{dz'}{(1+z')^{\alpha}}\frac{d\zL(z')}{dz'}\,.
$$}
\begin{equation}\label{varG}
\frac{d\rD}{dz}=\alpha (1+z)^{\alpha-1} \,\frac{\rM^0}{G_0}\
[G(z)-G(z^{*})]\,.
\end{equation}
where $z^{*}$ is the crossing point of the phantom divide, which
satisfies $\rD(z^{*})=\zL(z^{*})$. It follows that if $G$ is
asymptotically free -- that is to say, if $G(z)$ decreases with
redshift -- we should observe quintessence behavior
(${d\rD}/{dz}>0$) for $0\leqslant z\leqslant z^{*}$; whereas if $G$
is ``IR free'' (i.e. it increases with $z$, hence decreases with the
expansion), we should then have phantom behavior (${d\rD}/{dz}<0$)
in that interval. An interesting example of model of this kind
appears if we take once more the evolution law (\ref{CCH}) for the
vacuum energy,  but assume now that matter is
conserved\,\cite{SSS04}.  The basic cosmological equations can be
formulated in the following compact way:
\begin{eqnarray}\label{System1}
&&E^2(z)=g(z)\,\left[\OM(z)+\OL(z)\right]\ \ \ ({\rm Friedmann\ equation})\nonumber\\
&&(\OM+\OL)dg+g\,d\OL=0 \ \ \ \ \ \ \ \ \ ({\rm Bianchi\ identity})\label{bianchi2}\\
&& \OL(z)=\OLo+\nu\left[E^2(z)-1\right]\ \ \ \ \ \ ({\rm running\
\rL}) \nonumber\,, \label{RGlaw3}
\end{eqnarray}
where $E(z)\equiv H(z)/H_0$ and  $g(z)\equiv {G}(z)/{G_0}$. It is
now straightforward to solve them for the gravitational coupling,
which becomes a logarithmically running quantity of the Hubble rate:
\begin{equation}\label{GH} G(H)=\frac{G_0}{1+\nu\,\ln\left(H^2/H_0^2\right)}\,.
\end{equation}
We see that $G$ is asymptotically free (resp. IR free) for $\nu>0$
(resp. $\nu<0$). An alternative formulation of this running law for
$G$ and $\Lambda$, motivated by the quantum physics of inflation,
was presented in \cite{Fossil07}.

The previous considerations show that we have plenty of theoretical
and phenomenological motivations to support the possibility of
observing running cosmological parameters, specially through the
study of time-evolving vacuum cosmologies. Ultimately this issue
must of course be backed up by observations, and in this sense it is
very important to try to find all possible phenomenological
indications suggesting such a vacuum dynamical picture. It is also
remarkable that one could even find signs of the running $\CC$ in
the astrophysical context; for example, through the study of the
implications that a variable $\CC$ distributed (homogeneously or
inhomogeneously) in a large domain could have in the process of
local structure formation, see\,\cite{BPS10a}.

Time-evolving vacuum models inspired by QFT principles could do even
more: they could provide a possible solution to the cosmic
coincidence problem, and ultimately to the ``old cosmological
constant problem'', namely the CC fine-tuning problem -- the
toughest cosmological conundrum of all times\,\cite{CCproblem}.
Possible avenues leading to a significant step in this direction are
summarized in the next two sections.
\begin{figure}[t]
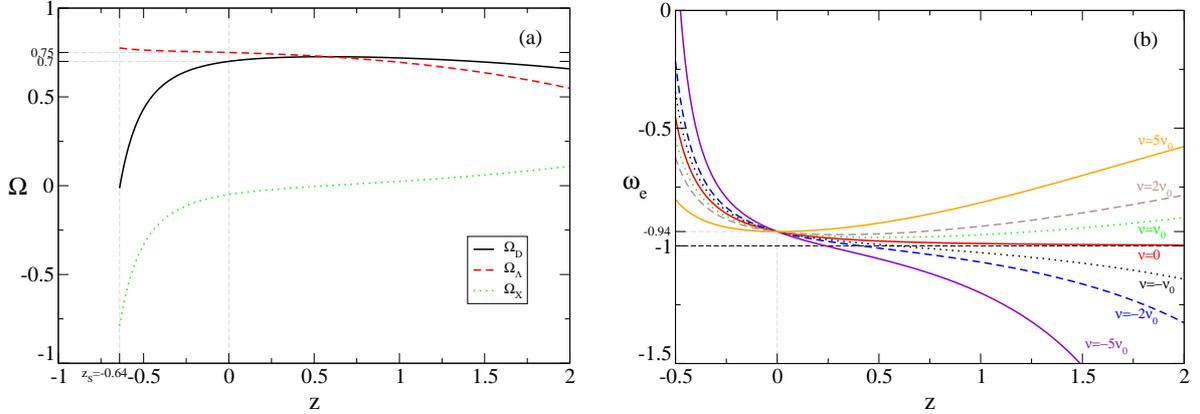

    \begin{tabular}{cc}\\
      \resizebox{0.47\textwidth}{!}{\includegraphics{irg4a.eps}}
      \hspace{0.3cm}
      \resizebox{0.47\textwidth}{!}{\includegraphics{irg4b.eps}} \\
    \end{tabular}
\caption{Features of the $\CC$XCDM model. \textbf{(a)} $\Omega (z)$
for the total DE and its individual components, $\Omega_X(z)$ and
$\OL(z)$, for $\OM^0=0.3,\, \OLo=0.75, \,\wX=-1.85,\,\,\nu=-\nu_0$.
The stopping of the expansion at $z=z_s$ is achieved here thanks to
$\Omega_X(z)$ becoming eventually negative; \textbf{(b)} Behavior of
the effective EOS, eq.(\ref{eEOS}), for the same $\wX$ and $\OLo$
and different values of $\nu$.}
  \label{fig4}
\end{figure}

\section{The $\CC$XCDM model: a possible solution to the cosmic coincidence problem }
\label{LXCDMmodel2}

The $\CC$XCDM model\,\cite{LXCDM12} makes a further step in our
systematic approach to a cosmology with variable cosmological
parameters. It is based on assuming that we have a composite DE
energy density $\rD=\rX+\rL$. In the canonical case where $G=const$,
it can be taken as covariantly conserved:
\begin{eqnarray}
\dot{\rho}_{\rm D}+
3(1+\we)\,\rD\,H=\dot{\rho}_{\Lambda}+\drX+\,3(1+\wX)\,\rX\,H=0\,,\label{fosc}
\end{eqnarray}
and in these conditions matter is self-conserved too of course:
$\dot{\rho}_{M}+\alpha\,H\,\rM=0$. Here $\rL=\rL(t)$ is the energy
density of the running $\CC$. Moreover, $\rX(t)$ is the energy
density of a new dynamical entity $X$ that we call the ``cosmon''.
Its dynamics is completely determined by the above local
conservation law (\ref{fosc}) once the evolution for $\rL(t)$ is
given. As emphasized in\,\cite{LXCDM12}, this implies that in
general it will be an effective entity, not a fundamental one (i.e.
$X$ is \textit{not} supposed to be a scalar field $\phi$) because
the dynamics imposed by (\ref{fosc}) will in general make it
incompatible with the scalar potential provided for $\phi$. The
cosmon $X$, instead, embodies the behavior of terms in the effective
action of QFT in curved space-time, and, far from being an
elementary scalar field, it represents a convenient way to
parametrize the effect of those terms in the cosmological evolution.
The EOS of the cosmon is indicated by $\wX$ (assumed, for the
moment, to be constant for simplicity). It should not be confused
with the effective EOS of the total DE, which reads
\begin{equation}\label{eEOS}
\we=\frac{p_D}{\rD}=\frac{-\rL+\wX\,\rX}{\rL+\rX}=
-1+(1+\wX)\,\frac{\rX}{\rD}\,.
\end{equation}
Next we will assume that $\rL=\rL(H)$ is the same function of $H$ as
in the running $\CC$CDM model presented in section \ref{RGmodel1} --
cf. eqs.\,(\ref{CCH}) and (\ref{C0C1}). Notice that this does not
mean that its redshift evolution is still given by eq.\,(\ref{CRG}),
as here the overall conservation law is no longer
(\ref{BianchiGeneral}) but (\ref{fosc}). After solving explicitly
the details of this new cosmological model -- which we refrain from
reporting here\,\cite{LXCDM12} -- the normalized energy densities
$\Omega (z)=\rho(z)/\rho_c^0$\,
($\rho=\rho_M,\rho_{\CC},\rho_X,\rD$) with respect to the present
critical density $\rho_c^0$ become determined as a function of the
redshift (cf. Fig.\,\ref{fig4}a). We see that the cosmon density
$\Omega_X$ can be negative, although this is not too surprising for
an effective quantity that does not represent a fundamental field.
On the other hand we observe that the total DE density
$\Omega_D=\OL+\Omega_X$ behaves properly, and its current value is
the measured DE value, i.e. we have $\Omega_D^0=0.7$ with
$\OM^0+\Omega_D^0=1$ (flat space). Furthermore, the EOS function
$\we=\we(z)$ can display a rich variety of behaviors while being in
agreement with the most recent data\,\cite{WMAP}. This is shown in
Fig.\,\ref{fig4}b, where we see that, depending on the value of
$\nu$, the EOS can be quintessence-like, or mimic (in some cases
almost exactly) a pure CC term ($\we\simeq-1$), and it can even
exhibit a mild transition from the phantom to the quintessence
regime. From Fig.\,\ref{fig4}b we recognize that this model is a
spectacular example of cosmology, in which despite of the fact that
the vacuum energy $\rL=\rL(H)$ is a running quantity, the effective
EOS of the composite DE can approach the standard $\CC$CDM model in
an arbitrary way, and may display only small ``quintessence-like''
or ``phantom-like'' departures from it, thus mimicking a scalar
field even in situations where such field could not behave
canonically ($\we\lesssim -1$).

\begin{figure}[t]
  \begin{center}
      \resizebox{0.60\textwidth}{!}{\includegraphics{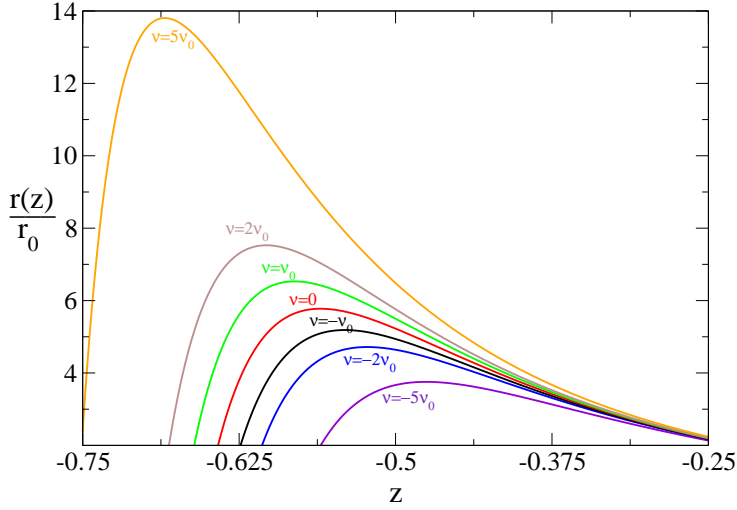}}
    \caption{Behavior of the ratio $r=\rD/\rM$, normalized to its current value $r_0$, for
$\wX=-1.85$, $\OMo=0.3,\OLo=0.75, \OXo=-0.05$ (flat space) and
different values of $\nu$. Here $\OXo<0$ insures stopping and
bouncing at a future point ($z=z_s<0$). The maxima satisfy $\,r_{\rm
max}<{\cal O}(10)$, which suggests a viable solution to the cosmic
coincidence problem.}
\end{center}
  \label{ratiorDrM}
\end{figure}
The truly nice feature of a model of this kind, however, is that it
can cure or highly alleviate the ``cosmic coincidence problem''.
Recall that this problem is related to the behavior of the ratio
$r(z)\equiv\rD(z)/\rM(z)$ between the dark energy density and the
matter energy density. In the standard $\CC$CDM model, that ratio
increases arbitrarily with the cosmic evolution because $\rD$ is
just the strictly constant vacuum energy density $\rL^0$, whereas
$\rM\to 0$ with the expansion, so there is no apparent reason why we
should find ourselves precisely in an epoch where $r={\cal O}(1)$.
Or, to put in other terms: there is no obvious reason why the epoch
where the universe started to accelerate is so recent. This is an
unexplained coincidence puzzle in the standard model. Not so in the
$\CC$XCDM model, in which, thanks to the presence of the cosmon $X$,
the function $r=r(z)$ presents a maximum at the resdhift
\,\cite{LXCDM12}:
\begin{equation}\label{zs}
z_{\rm max}=\left[\frac{\OL^0-\nu}{\wX\,(\Omega_X^0+\nu\,\OM^0)-
\epsilon\,(1-\OL^0)}\right]^{\frac{1}{3\,(1+\wX-\epsilon)}}-1\,,
\end{equation}
where we have defined $\epsilon\equiv\nu(1+\wX)$. One can show that
this point cannot be in our past\,\cite{LXCDM12}, so it will
generally be in the future (hence $-1<z_{\rm max}<0$). Beyond that
maximum, there is a turning point $z=z_s<z_{\rm max}$ where the
universe stops and reverses its evolution. At the stopping point,
$\Omega_D(z_s)=-\OM(z_s)<0$. This is possible because we can have
$\Omega_X<0$, or $\OL<0$.

The following observation is now in order: notice, from
eq.\,(\ref{fosc}), that for $\nu=0$ (which implies no running of
$\rL$) the cosmon energy density is conserved; so, if in addition
$\Omega_X^0=0$ (vanishing current cosmon density), we must then have
$\Omega_X(z)=0$ at all times. Therefore, in the limit
$\nu=\Omega_X^0=0$ we exactly recover the standard $\CC$CDM model.
And it is precisely in this limit that the formula (\ref{zs}) for
the redshift location of the maximum of the ratio $r(z)$ ceases to
make sense, as it is evident by simple inspection of that formula.
In contrast, when we make allowance for the parameter space of the
$\CC$XCDM model, the maximum does exist and moreover the ratio
$r(z)$ stays bounded -- typically $r<{\cal O}(10)$ -- for the entire
history of the universe. This is clearly seen in Fig.\,3, for
various values of the parameters. So, in fact, the $\CC$XCDM model
can provide a nice solution to the cosmic coincidence problem
without necessarily departing from the $\CC$CDM model in a
significant way. Furthermore, the detailed analysis of the cosmic
perturbations in the $\CC$XCDM model indicate excellent
compatibility with the present observational data on structure
formation\,\cite{Grande08}.

\section{The ``new cosmon'' model: towards solving the ``old CC problem'' }
\label{RelaxedUniverse}

The ``Relaxed Universe'' is a special realization of the $\CC$XCDM
model which points to a more ambitious aim, to wit: solving the
``old CC problem''\,\cite{CCproblem}, i.e. understanding how the
Universe could start up its thermal history with a huge value of the
initial vacuum energy $\rLi\sim M_X^4$ (e.g. triggered by a GUT
with, say $M_X\sim 10^{16}$ GeV, presumably responsible for
inflation) and then eventually ended up with its current tiny value
$\rL^0\sim 10^{-47}$ GeV$^4\sim 10^{-110}\rLi$\,! The history of
this profound theoretical conundrum, which keeps in check the
fundamental physics principles since more than forty years ago,
traces back to Zeldovich's realization in the late
sixties\,\cite{Zeldovich67}, that the vacuum fluctuations of the
quantum fields should induce a very large value of the CC. Since
then the problem stays with us, unbeaten and virtually invincible.
The ``traditional'' fine-tuning solution \,\cite{CCproblem} (i.e.
the sheer adjustment by hand of the initial big CC value $\rLi$ by a
carefully chosen finite ``counterterm'') is simply and fully
inadmissible, especially at the quantum level -- see
\cite{BSSFull10a} (particularly Appendix B of this reference) for a
detailed account. A much more promising approach, instead, should be
to find out a mechanism that neutralizes \textit{dynamically} the
presence of the initial vacuum energy, $\rLi$, whatever it be its
seed value in the early universe. We will call the cosmological
model related to this relaxation mechanism, the ``Relaxed
Universe''\,\cite{BSSFull10a}\footnote{For other recent alternative
approaches aiming to palliate the CC problem, see e.g.
\cite{alter}}.

As we would like that the expansion of the universe be responsible
for the triggering of the relaxation dynamics, it is reasonable to
assume that such mechanism should be closely related to the
expansion itself, or more precisely, to the interaction governing
the expansion, i.e.\ {\em gravity}. A note of caution, however:
although we aim at a modified form of gravity, we cannot just
content ourselves with a ``late time effect'' (as usually done in
the literature on modified
gravity~\cite{Capozziello:2003tk,Carroll:2003wy,Odintsov06}). We
rather must have a \textit{persistent} effect, fully active
throughout the \textit{entire history} of the universe. Indeed, in
order to solve the old CC problem we need that some automatic
mechanism sets to work at a time prior to the onset of the radiation
epoch -- at which time it must cancel most of the value of the
original vacuum energy present in the early universe-- \textit{and},
most important, we also need that this cancelation is maintained for
the rest of the universe lifetime, leaving a small value today and
maybe also in the future.

It is not easy to device such a mechanism, but some ideas on it are
briefly explained here -- see \cite{BSS0910}, and specially
\cite{BSSFull10a}, for a detailed account; and
\cite{Bauer0910,Stefancic08} for alternative forms. The effective
action of the ``Relaxed Universe'' is proposed in the following way:
\begin{equation} \mathcal{S}=\int
d^{4}x\,\sqrt{|g|}\left[\frac{1}{16\pi G}R-\rLi-
\F(R,\G)+\mathcal{L}_{\varphi}({\rm matter\
fields})\right]\,,\label{eq:CC-Relax-action}\end{equation}
in which we have an (arbitrarily large) initial vacuum energy
$\rLi$, and we have included a functional (the ``cosmon
functional'') $\F=\F(R,\G)$ of the Ricci scalar $R$ and the
Gauss-Bonnet invariant
$\G=R^{2}-R_{\mu\nu}R^{\mu\nu}+R_{\mu\nu\rho\sigma}R^{\mu\nu\rho\sigma}$.
The structure of this functional is convenient in order to avoid the
presence of ghost degrees of freedom and also to dodge the
Ostrogradski instability, i.e.\ the appearance of vacuum states of
negative energy\,\cite{Woodard07}.  The corresponding (generalized)
Einstein equations read
\begin{equation}\label{eq:Mod-Einstein-Eqs}
R_{\,\,\nu}^{\mu}-\frac12\,\delta^{\mu}_{\nu}\,R= -8\pi
G\,\left[\rLi\,\delta_{\,\,\nu}^{\mu}+2
E_{\,\,\nu}^{\mu}+T_{\,\,\nu}^{\mu}\right]\,,
\end{equation}
where $T_{\,\,\nu}^{\mu}$ is the ordinary energy-momentum tensor of
the matter fields, and $E_{\,\,\nu}^{\mu}$ is the new tensor in the
field equations associated to the presence of the cosmon functional
$\F$ in the action. We assume that matter is covariantly conserved,
and therefore $E_{\,\,\nu}^{\mu}$ must also be conserved:
$\nabla^{\nu}E_{\,\,\nu}^{\mu}=0$. Clearly, our functional
constitutes an extension of the Einstein-Hilbert's action with CC,
in which we have added the $\F(R,\G)$ part. The next point is to
make an appropriate choice for the functional $\F$ such that it can
have a bearing on the cosmological constant problem. We make the
following Ansatz:
\begin{eqnarray}\label{eq:FFP}
\F(R,\G)&=&\beta\, F(R,\G)=\frac{\beta}{B(R,\G)}\,,\\
B(R,\G)&=&\frac{2}{3}R^{2}+\frac{1}{2}\G+(y\,
R)^{3}\,,\label{eq:defB}
\end{eqnarray}
where $B(R,\G)$ is a polynomial of the Ricci scalar and the
Gau\ss-Bonnet invariant, and $(\beta, y)$ are dimensionful
parameters to be determined later. This model is in fact of
$\CC$XCDM type, in the sense that the cosmon $X$ here is related to
the energy density $\rF= 2 E_{\,\,0}^{0}$ generated by $\F$ in the
effective action, and so $\rF$ plays the role of $\rX$ in the
$\CC$XCDM model. Explicitly, one finds
\begin{eqnarray}
\rF=2\,E_{\,\,0}^{0} = {\beta}\left[ F(R,\G) -6(\dot{H}+H^{2})F^{R}
+6H\dot{F}^{R} -24H^{2}(\dot{H}+H^{2})F^{\G} +24H^{3}\dot{F}^{\G}
   \right]\,.\label{eq:E00}
\end{eqnarray}
In our case, $F$ has the form $F=1/B$ as defined in (\ref{eq:FFP}),
and  $F^{Y}\equiv\partial F/\partial Y$ are the partial derivatives
of $F=F(R,\G)$ with respect to~$Y=R,\G$. The parallelism with the
$\CC$XCDM model becomes even more transparent if we notice that the
mixed DE energy density defined in (\ref{fosc}) takes here the form
$\rLe(t)=\rLi+\rF(t)$, and is covariantly conserved. Indeed, since
$\rLi$ is constant, we have
$\dot{\rho}_F+3\,H\,(1+\omega_F)\,\rF=0$, where $\omega_F$ is the
effective EOS of the cosmon, i.e. the analog of $\wX$ in section
\ref{LXCDMmodel2}, except that here $\omega_F=\omega_F(t)$ is a
non-trivial function of the cosmological evolution which is
determined by the structure of the cosmon functional (\ref{eq:FFP}).

This is indeed a good example of the situation we have mentioned in
section \ref{LXCDMmodel2}, in which we emphasized that the cosmon
$X$ is \textit{not} a fundamental field, but a non-trivial
field-theoretical object generated from terms of the effective
action of QFT in curved space-time. Evaluating it in the FLRW metric
we obtain, for its denominator (\ref{eq:defB}):
\begin{equation}
B=24H^{4}(q-\frac{1}{2})(q-2)+\left[6\,y\,H^{2}(1-q)\right]^{3}\,,\label{eq:B}
\end{equation}
where $q=-1-\dot{H}/H^2$ is the deceleration parameter. The previous
formula shows why the cosmon functional (\ref{eq:FFP}) can provide a
dynamical mechanism for counterbalancing the value of $\rLi$,
whatever it be its size. The relaxation of the effective vacuum
energy $\rLe(t)=\rLi+\rF(t)$ in the radiation epoch originates from
the~$(1-q)$ factor in the denominator (\ref{eq:B}) of the cosmon
functional, as this term grows as $\sim H^6$ at large $H$ and
dominates $B$ in the earlier epochs. The large~$\rLi$ left over in
the immediate post-inflationary period drives the deceleration
parameter~$q$ to larger values until $q\to 1$, which corresponds to
radiation-like expansion. In other words, the very existence of the
radiation period is triggered automatically by the presence of this
term, which can be thought of as a countermeasure launched by the
universe against the presence of the large ``residual'' vacuum
energy $\rLi$ in the pre-radiation era. As a result, $\rF= 2
E_{\,\,0}^{0}$ is driven \textit{dynamically} to very large values,
opposite in sign to $\rLi$, and produce a small effective $\rLe$ in
the radiation epoch, for an appropriate sign of the parameter
$\beta$. Similarly, the $(q-1/2)$ factor in the denominator
(\ref{eq:B}) will be responsible for the relaxation in the matter
era~($q\simeq {1}/{2}$), at lower values of $H$. This is possible
because the term with the $(q-1/2)$ factor is weighted by a power of
$H$ smaller than the term carrying the $(1-q)$ factor. The
transition between the two regimes (i.e. from radiation to matter)
is what fixes the parameter $y\sim H_{\rm eq}^{-2/3}$, where $H_{\rm
eq}\sim 10^{5}\, H_{0}$ is the Hubble rate just at the transition
time between the two epochs.

\begin{figure}[t]
  \begin{center}
      \resizebox{1.00\textwidth}{!}{\includegraphics{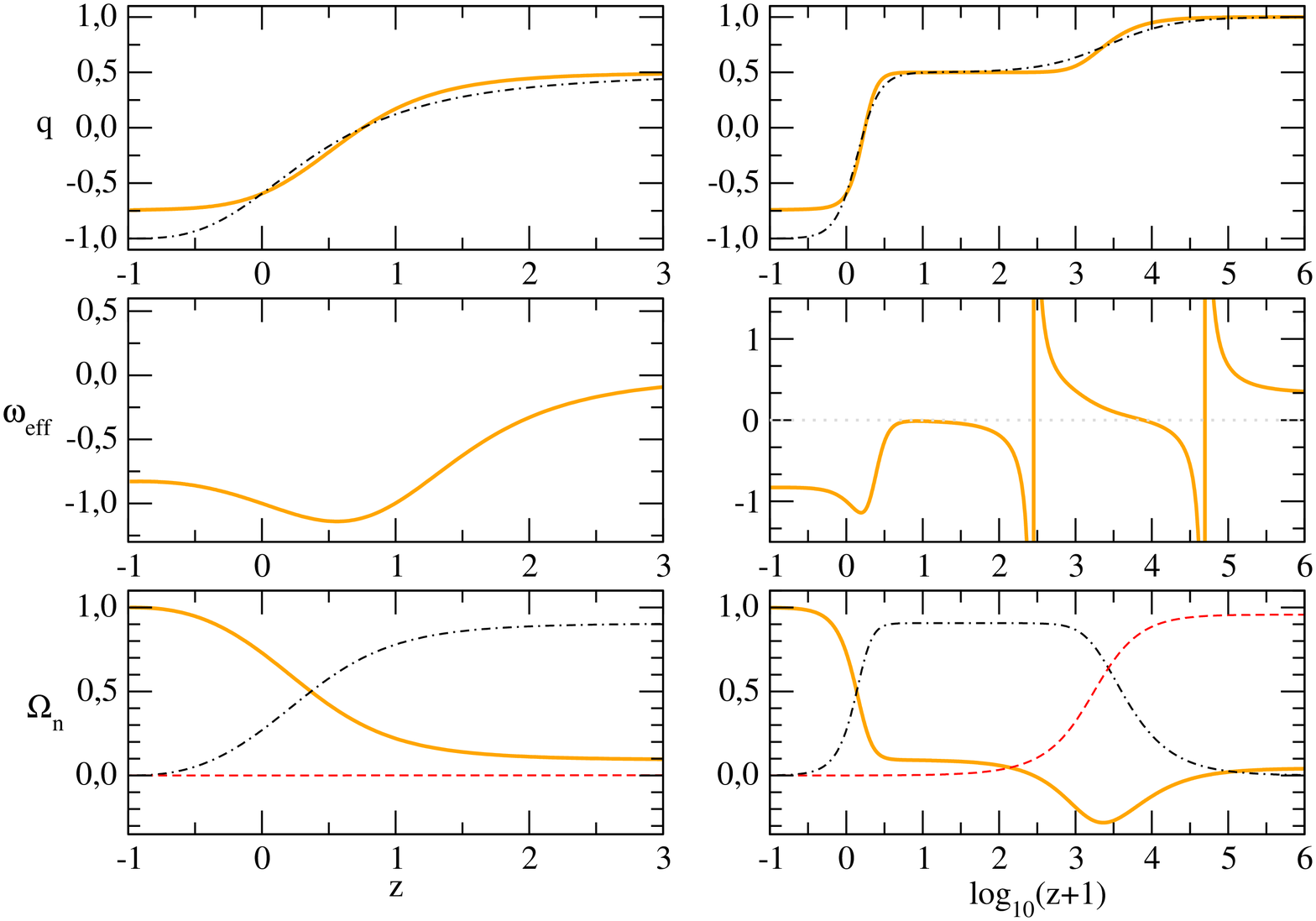}}
    \caption{Deceleration parameter~$q$, effective EOS~$\weff$, and
relative energy densities for each energy component (normalized with
respect to the corresponding critical density)
~$\Omega_{n}(z)=\rho_{n}(z)/\rho_{c}(z)$ of dark energy~$\rLe$
(orange thick curve), dark matter~$\rho_{M}$ (black dashed-dotted)
and radiation~$\rho_{r}$ (red dashed) as functions of redshift~$z$
in the relaxation model (\ref{eq:FFP}) with $y=0.7\times
10^{-3}\,H_0^{-2/3}$, $\rLi=-10^{60}$ {GeV}$^4$,
$\Omega_{M}^{0}=0.27$, $\Omega_{r}^{0}=10^{-4}$, $q_{0}\approx-0.6$,
$\dot{q}_{0}=-0.5\,H_0$. In the $q$-plot, the thick orange curve
corresponds to the Relaxed Universe, and the black dashed-dotted
curve to $\Lambda$CDM.}
  \label{RelaxHistory}
  \end{center}
\end{figure}

A numerical example illustrating all these features can be
appreciated in Fig. \ref{RelaxHistory}. The Relaxed Universe is seen
to follow very closely the past history of the standard $\CC$CDM
model (see especially the upper panel in that figure), and its
effective EOS loiters around the $\CC$CDM value $\weff\simeq -1$
near our time (see the second panel of that figure). In the far
past, $\weff$ turns out to track the values $\weff\sim 0$ and
$\weff\sim 1/3$ in the matter and radiation epochs, respectively.
%
%
Finally, when the universe leaves the matter epoch $q\simeq 1/2$,
there is a kind of impasse, as there is no longer a special value
for $q$ to pick dynamically in order to maintain the counterbalance
of the large $\rLi$ by the cosmon density $\rF$. What to do now?
Under these circumstances, the last resort of the relaxation
mechanism is to enforce (dynamically) a very low value of $H$. Since
in the current universe we must have $|\rLi+\rho_{F}(H)|/\rLi\ll 1$,
with $\rF\sim\beta/H^4$, it follows that the value of $H$ that
solves this equation is approximately given by
\begin{equation}\label{Hstar}
H_{*}\sim \left(\frac{\beta}{|\rLi|}\right)^{1/4}\,.
\end{equation}
For an appropriate choice of $\beta$, we may bring $H_{*}$ close
enough to the current value $H_0$. As $\beta$ has dimension $8$ of
mass, we can rewrite it as $\beta\equiv M^8$. The relaxation
condition requires $\rF^0\simeq -\rLi$ where $|\rLi|\sim M_X^4$.
Thus, $M^{8}\sim |\rF^0|\,H_0^4\sim M_X^4\,H_0^4$. Taking the
standard GUT scale $M_X\sim 10^{16}$ GeV, we obtain ${M}$ in the
ballpark of a light neutrino mass:
\begin{equation}\label{Mcanonical}
{M}\sim \sqrt{M_X\,H_0}\sim 10^{-4}\,{\rm eV}\,.
\end{equation}
Incidentally, this value for $M$ is the geometric mean of the two
most extreme mass scales available in our universe below the Planck
mass. It is rewarding to see that we do not obtain an extremely
small mass scale of order $m_{\phi}\sim H_0\sim 10^{-33}$ eV, as in
the case of quintessence models.

From the above considerations, we see that the $F$-functional's main
mission is to \textit{dynamically} counterbalance the big initial
$\rLi$ at \textit{all epochs}, not just now; and in the last stage
of this dynamical process the Hubble rate is driven to the very
small value $H_0$ that we observe today. At the end of the day, what
we get is not just a late time effect of the cosmological evolution
(as in the traditional modified gravity models), but rather a truly
perennial large scale influence on all periods of the cosmic
history. One hope is that a relaxation mechanism of this sort should
ultimate provide a dynamical solution to the tough fine tuning
problem underlying the ``old CC problem''\,\cite{CCproblem}. For
example, in the standard model of electroweak interactions such
fine-tuning involves $55$ decimal places, at least, and it must be
retuned order by order in perturbation theory until reaching loop
diagrams of order 20th! --see \cite{BSSFull10a} for details. In the
Relaxed Universe, in contrast, the tuning to all orders is automatic
and performed dynamically by the universe's expansion. This is why
such field-theoretical object associated to the functional $F$ is
called the ``cosmon'', as its primary mission is to compensate for
the huge initial CC throughout the entire cosmological evolution.
Clearly, it has the same aim as the cosmon scalar field first
introduced in \cite{PSW}, except that here the cosmon has nothing to
do with scalar fields, and moreover is \textit{not} subdued by
Weiberg's ``no-go'' theorem\,\cite{{CCproblem}} --
see\,\cite{BSSFull10a} for a more detailed account.

While in the present version the cosmon is not a fundamental
quintessence field\,\cite{Quintessence}, it can perfectly mimic
quintessence-like behavior (cf. Fig. \ref{RelaxHistory}). The new
cosmon, however, is a more complex object: it is a functional whose
ultimate origin must be provided by a fundamental theory (say string
theory or any other theory aiming at a final explanation of quantum
gravity). In the meanwhile, what we have shown here is a practical
working example, actually a class of such functionals which are able
to truly adjust the vacuum energy dynamically, whatever it be its
initial value, and without performing fine-tuning at any stage of
the cosmological evolution. As this was the main aim of the original
cosmon model\,\cite{PSW}, it is naturally suggested that the general
class of these functionals is called the ``new cosmon''. For more
details, see \cite{BSSFull10a}.

Let us also mention that the Relaxed Universe based on the cosmon
functional (\ref{eq:FFP}) and extensions thereof (cf.
Ref.\cite{BSSFull10a}), provides also a possible natural solution to
the cosmic coincidence problem (mentioned in
section\,\ref{LXCDMmodel2}). This additional bonus is not too
surprising if we recall that the Relaxed Universe is a particular
implementation of the class of $\CC$XCDM models of the cosmic
evolution\,\cite{LXCDM12}. We have already mentioned that, in the
Relaxed Universe, there is a tracking behavior of matter/radiation
and DE during the $q=1$ (radiation) and $q=1/2$ (matter epochs).
However, when the matter epoch is left behind and the relaxation
mechanism is enforced (dynamically) to pick a very small value of
$H$ (as we have seen above) in order to insure the condition
$|\rLi+\rho_{F}(H)|\ll\rLi$ to hold, this crucial event decoupled
once and forever the tracking behaviors of matter/radiation and DE,
and since then the DE dominates the universe's expansion. Such
breakdown of the tracking property at the end of the matter epoch,
on the other hand, pushed the subdominant DE density curve upwards
until crossing the decaying DM one at some point after the matter
epoch and hence near our time. This feature, which can be very well
appreciated in Fig. \ref{RelaxHistory} (cf. third panel), encodes
the very origin and the possible exciting explanation (within the
Relaxed Universe) of the ``coincidence $\OL^0\sim\OM^0$'' of the
matter and DE densities in our late time neighborhood. Not only so:
we can also understand now the smallness of both $\rLo$ and $H_0$ in
front of the huge initial vacuum energy density $\rLi$. Indeed, from
(\ref{Hstar}) we predict $H_0\sim M^2/M_X\sim 10^{-42}$ GeV, for
${M}$ not far from the $\sim$meV scale, which is the mass scale
itself of the current CC: $m_{\CC}\equiv
\left(\rL^0\right)^{1/4}\sim 10^{-3}$ eV. We have thus arrived at a
most natural scenario (``The Relaxed Universe'') in which, at
variance with the quintessence formulation of the DE, there is no
need to severely fine-tune any parameter of the model, and moreover
it is not required the existence of any extremely tiny mass scale.

\section{Conclusions}
We have shown that models with variable vacuum energy
$\rho_{\Lambda}$ (and may be also with variable $G$) generally lead
to a non-trivial effective EOS when viewed from the point of view of
a self-conserved DE fluid at fixed $G$. As a result, these models
can effectively appear as quintessence, and even as phantom energy,
without need of invoking the existence of fundamental quintessence
and phantom fields. Combining a time-evolving $\rho_{\Lambda}$ with
a variable ``cosmon'' energy density $\rX$ (in general \textit{not}
representing any scalar field), we obtain the class of $\CC$XCDM
models, which are able to deal with the cosmic coincidence problem.
Finally, one can provide a version of the $\CC$XCDM model in which
the new cosmon stands for a self-adjusting functional, namely one
which is able to counteract the effects of an arbitrarily large CC
in the early universe, leaving a small and completely innocuous net
value of the CC at any stage of the cosmological evolution, in
particular the tiny value $\rLo\sim 10^{-47}$ GeV$^4$ that we
observe at present. This particular cosmon functional is thus able
to reduce the value of the CC \textit{without fine-tuning}, and may
be eligible as a prototype model (``The Relaxed Universe'') for
eventually solving the old CC problem.

\ack I have been partially supported by DIUE/CUR Generalitat de
Catalunya under project 2009SGR502 and by MEC and FEDER under
project FPA2007-66665, and also by the Consolider-Ingenio 2010
program CPAN CSD2007-00042. It is my pleasure to thank  S.
Basilakos, F. Bauer, J.C. Fabris, J. Grande, A. Pelinson, M.
Plionis, I.L. Shapiro and H. \v{S}tefan\v{c}i\'{c} for a nice and
stimulating collaboration in some of the topics discussed here. I
would also like to thank  R. Woodard, B.F.L. Ward, M. Maggiore and
N. Bili\'c for discussions. Last but not least, I am also very
grateful to L. Perivolaropoulos and all the organizers of NEB 14,
for the invitation to this excellent hellenic conference on Recent
Developments in Gravity.

\end{document}